\input epsf
\newwrite\ffile\global\newcount\figno \global\figno=1

\def\writedef#1{}

\def\figin{\epsfcheck\figin}\def\figins{\epsfcheck\figins}
\def\epsfcheck{\ifx\epsfbox\UnDeFiNeD
\message{(NO epsf.tex, FIGURES WILL BE IGNORED)}
\gdef\figin##1{\vskip2in}\gdef\figins##1{\hskip.5in}
\else\message{(FIGURES WILL BE INCLUDED)}%
\gdef\figin##1{##1}\gdef\figins##1{##1}\fi}

\def\figinsert{}
\def\ifig#1#2#3{\xdef#1{fig.~\the\figno}
\writedef{#1\leftbracket fig.\noexpand~\the\figno}%
\figinsert\figin{\centerline{#3}}\medskip\centerline{\vbox{\baselineskip12pt
\advance\hsize by -1truein\center\footnotesize{  Fig.~\the\figno.} #2}}
\bigskip\endinsert\global\advance\figno by1}
\def\endinsert{}

\documentclass{ws-p9-75x6-50}

\begin{document}

\title{Renormalization Group Flows In The Deformed AdS/CFT Correspondence}

\author{Nick Evans}  

\address{Department of Physics and Astronomy, University of Southampton,
                Highfield, Southampton, SO17 1BJ, UK}

\maketitle

\abstracts{In the AdS/CFT correspondence motion in the radial 
direction of the AdS space is identified with renormalization
group flow in the field theory. For the N=4 Yang-Mills theory this 
motion is trivial. More interesting examples of renormalization
group flow occur when the N=4 theory is deformed. Aspects of the
flows are discussed for the N=4 theory on its moduli space, and
deformed to N=2 in the infra-red within the context of 5d SUGRA.
10d lifts and brane probing are crucial tools
for linking the spacetime backgrounds to the dual field theory.
}

\section{Introduction}\vspace{-0.1cm}

There is now a well established and remarkable correspondence between
N=4 super Yang-Mills (SYM) gauge theory in 3+1 dimensions and type IIB
string theory on $AdS_5 \times S^5$ spacetime\cite{mald,review}. 
The renormalization group
(RG) appears to be an  integral component of the duality with changes
in RG scale in the gauge theory 
corresponding to motion in the ``holographic'' radial
direction of the $AdS_5$ space on the gravity side\cite{RG}. 
As will be reviewed 
below this role for RG flow was deduced in the original case of the
N=4 SYM theory. The RG flow in this theory is trivial however since
the theory is conformal. There has therefore been considerable
interest in studying more complicated examples of the correspondence where the 
RG flow is more dynamic. A sensible way in which to construct such
theories is to begin with the N=4 conjecture and use the symmetries
on each side of the duality to understand how to include relevant
deformations\cite{def}. This article will attempt to provide a brief overview
of the state of such investigations. To maximally simplify the effort 
on the string theory side of the duality initial studies were performed
in the 5 dimensional supergravity (SUGRA) theory describing the string 
theory at long distances relative to the $S^5/AdS_5$ radius. As will
emerge though the results have lead back to elements (such as branes)
of the full string theory requiring more complex analysis. In addition
the deformed SUGRA backgrounds discovered need considerable work 
to extract/confirm properties of their dual field theories. In this respect
brane probing has been found to be a sharp tool. As examples of 
non-conformal theories I will discuss the N=4 theory on moduli space 
as investigated in\cite{FGPW} 
and the N=4 theory deformed by the inclusion of
a mass term to N=2\cite{n2,PW,n21}. Much of my discussion is 
based on work with M. Petrini and C.V. Johnson. \vspace{-0.2cm}

\section{The AdS/CFT Correspondence}

The $AdS$/CFT correspondence is between N=4 SYM in 3+1 dimensions
(a theory with four adjoint spinors, 6 real scalars and an SU(N) gauge group)
and Type IIB string theory on the space $AdS_5 \times S^5$. The $AdS$
space is given by the metric
\begin{equation}
ds^2 = {du^2 \over u^2} + u^2 dx^\mu dx_\mu
\end{equation}
and may therefore be thought of as a stack of 3+1 dimensional 
Minkowski spaces with a warp factor in the perpendicular $u$ direction.
As the correspondence was first proposed the N=4 SYM theory lived
on the top most such Minkowski sheet at $u = \infty$. 
The bulk string theory at long distances 
may be replaced by Type IIB supergravity. The N=4 SYM theory and the bulk
states therefore naturally interact at the boundary. We allow all
action terms of the form
\begin{equation}
S = \int dx^4  {\cal O}_{FT} \Phi_0
\end{equation}
where ${\cal O}_{FT}$ is a gauge invariant operator in the field theory
and $\Phi_0$ is the boundary value of some supergravity
field. Of course only terms that respect the global symmetries of each
theory are allowed as we will discuss shortly. The vevs of supergravity fields 
at the boundary act as source terms in the field theory.
It is immediately apparent
how one would go about introducing conformal symmetry breaking operators
into the boundary theory. Operators such as ${\cal O}_{FT} J$ can give
rise to mass terms and operators of the form $|{\cal O}_{FT}- J|^2$ 
can be used to force the expectation values of fields to
explore the moduli space of a theory.

The full statement of the duality is that the action of the classical 
(weakly coupled) supergravity theory evaluated with appropriate boundary
conditions on the fields is the full generating functional for the quantum
N=4 SYM theory (at large 'tHooft coupling $g^2N$) with the corresponding 
source terms. 

The most immediate check is that the global symmetries of the two theories
match. The N=4 SYM theory has a SU(4)$_R$ global symmetry. This is matched by 
the isometry group of the $S^5$ on the supergravity side of the duality.
The N=4 SYM is also conformal thus having a super-conformal symmetry group 
SO(2,4) - for example there are dilatations of the form
\begin{equation}
x \rightarrow e^\alpha x, \hspace{1cm} \phi \rightarrow e^{- \alpha} \phi
\end{equation}
that leave the massless scalar action invariant
\begin{equation}
\int dx^4  \partial^\mu \phi \partial_\mu \phi
\end{equation}
The supergravity theory is not conformal though (there are many mass
scales associated with different Kaluza Klein states on the $S^5$) so this 
symmetry must be realized as a pure spacetime symmetry. The transformation on 
$x$ in (3) can be made a symmetry of the line element in (1) provided
\begin{equation}
u \rightarrow e^{- \alpha} u
\end{equation}
The full SO(2,4) symmetry can also be realized as a spacetime symmetry.

We have just seen that the duality implies that the radial $u$ 
direction of $AdS$ transforms as a mass dimension under dilatations in 
the field theory. The natural correspondence to make for motion in
$u$ is therefore change of RG scale in the field theory! We now let the
field theory live on any of the Minkowski slices of $AdS_5$ but the action
defined there by the supergravity field vevs is that of the field theory 
at a different RG scale. For the gravity dual to the N=4 SYM at the
origin of moduli space there are no sources switched on and hence
no supergravity field vevs. The supergravity theory looks the same on 
every Minkowski space slice through $AdS_5$. This, of course, is a result
of the N=4 theory being conformal. Let us now think about some 
examples of related theories where the conformal symmetry is broken by the 
introduction of relevant operators.\vspace{-0.2cm}

\section{N=4 SYM on Moduli Space}

The moduli space of the N=4 SYM theory is characterized by a vev for
the gauge invariant quantity $tr \phi^2$ so we must include a source term
$|tr \phi^2 - J|^2$. There are six real scalar fields transforming as the 6 of
SU(4)$_R$ so $tr \phi^2$ is in general a 6x6 matrix. The symmetric piece
is the 20 representation of SU(4)$_R$ so we can look for a source in the 20.
The simplist supergravity analysis is to use the 5 dimensional truncation
of Type IIB supergravity on AdS$_5 \times S^5$ which is valid on scales 
much larger than the radius of curvature and consists of only the lightest
Kaluza Klein states. One of these is a scalar field in the 20 representation,
$\alpha$.
Even the truncated theory is quite complicated with the scalar fields
transforming under the coset space $E_6/USp(8)$. An expression for the 
potential was though calculated in \cite{gun} in the 1980s and we can proceed
with those results. A notable feature of the potentials is that they
are unbounded from below! In fact this is not as pathological as it 
first seems since the energy of the scalars receives a contribution 
from the positive curvature of the spacetime. This contribution allows
a limited negative curvature. 

As a sample flow consider turning on a vev  $tr \phi^2 = diag(1,1,1,1,-2,-2)$
\cite{FGPW}
which corresponds to considering a particular element of the 20. We can
look for non-zero solutions of the equation of motion of the appropriate scalar
in the supergravity. The scalar vev in the supergravity setting 
though can give rise to a backreaction on the metric. We will only
consider variation in the $u$ direction so we can parametrize the metric as
(writing $u = e^y$)
\begin{equation}
ds^2 = dy^2 + e^{2 A(y)} dx^\mu dx_\mu
\end{equation}
The equations of motion for the scalar $\rho = e^\alpha$ and $A$ are
\begin{equation}
{\partial \rho \over \partial y} =- {2 \over \sqrt{6}} \left(
\rho^5 - {1 \over \rho} \right), \hspace{1cm} 
{\partial A \over \partial y} =  {1 \over 3} 
\left( \rho^4 + {1 \over 2 \rho^2}\right)   
\end{equation}
In Fig 1 some sample solutions are shown. For large $y$ (the UV) 
$\phi \sim y$, the $AdS$ limit,  and $\rho \sim exp(2 y)$, showing
that the deformation indeed has dimension 2. For small $y$ 
$\rho$ grows (as the scalar vev rolls down the unbounded potential)
in a way reminiscent of the RG flow of a dimension 2 operator in the
field theory.
A worrying aspect of the solutions is that the spacetime becomes singular
at some finite value of $y$. Are these spacetimes good solutions? 
The singularities are an indication that the SUGRA theory is not 
the full story. The $AdS$/CFT duality is with the $\alpha' \rightarrow 0$
limit of the full closed string theory and in particular we can expect to
find D-brane configurations occuring. The SUGRA theory does not
know about D-branes so will become singular where its knowledge fails.
Remember that the correspondence was 
deduced by constructing the N=4 gauge theory as the low energy description
of open strings living on the surface of a coincident stack of D3 branes at
$y = - \infty$. To place the theory on its moduli space
we would separate the D-branes. In \cite{FGPW} it was shown that this metric 
is indeed that describing the SUGRA background about 
such a D-brane distribution. These are therefore physical solutions.

\begin{figure}
\epsfxsize 12cm \centerline
{\epsffile{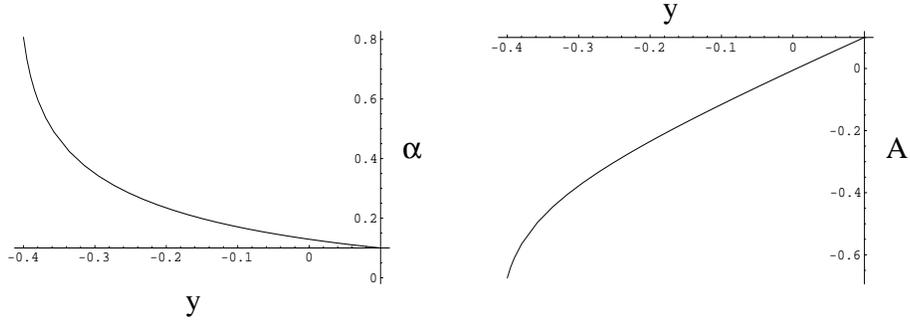}}
\caption{Sample solutions of the 5d SUGRA scalars for the N=4 theory
on moduli space.
}
\end{figure}

So far we have not shown any hard evidence for the claimed duality
between these spacetimes and the N=4 theory on moduli space. To do so we
need a concrete tool that makes a connection between the two descriptions.
Such a tool is brane probing; we imagine separating one of the D3 branes 
from the continuous distribution and move it about in the spacetime.
The low energy description of the D3 brane is given by the Born-Infeld
action and describes the U(1) gauge theory from the open string spectrum
on its surface. The important point is that the {\it scalar fields} in the
field theory on the brane correspond to the {\it position} of the brane in
the spacetime. The action is
\begin{equation}
S = - \mu \int d^4 \xi {\rm det} ^{1/2} \left[ G_{ab}  + e^{-\phi/2}  F_{ab}
\right]  + \mu \int C^4
\end{equation}
where $\mu$ is the D3 tension, $\phi$ the dilaton, $F^{ab}$ the gauge
potential on the brane, $C^4$ the Ramond Ramond 4-form potential to
which a D3 couples and finally $G^{ab}$ is the pull back of the spacetime
metric
\begin{equation}
G^{ab} = G_{\mu \nu} { \partial x^\mu \over \partial \xi^a} 
{ \partial x^\nu \over \partial \xi^b}
\end{equation}
The probe therefore provides the link between the background spacetime 
metric and the low energy field theory variables. 

Of course to perform the probe calculation we need the full 10D SUGRA
solution rather than just the 5D solution. The use of the 5D truncation though
is that for any solution 
there is automatically a lift to 10D though one requires
considerable technology to perform the lift. Thankfully Pilch and 
Warner\cite{PW} are 
experts in such matters and have provided us with the lift:
\begin{equation}
ds^2 = {X^{1/2} \over \rho} ( e^{2 A} dx^2 + dr^2) + {X^{1/2} \over \rho^3}
( d \theta^2 + {\rho^6 \cos^2 \theta \over X} d \Omega_3^2 + {\sin^2 \theta
\over X} d \phi^2 )
\end{equation}
where
\begin{equation}
X = \cos^2 \theta + \rho^6 \sin^2 \theta
\end{equation}
In addition the solution has a non-trivial 4-from potential
\begin{equation}
C^4 = { e^{4A} X \over \rho^2} dx^0 \wedge dx^1 \wedge dx^2 \wedge 
dx^3 \wedge 
\end{equation}

We may now perform the brane probe by substituting this background
spacetime into the Born Infeld action. For a static probe we find
the action vanishes exactly no matter the position of the brane in the 6 
transverse dimensions.
In the U(1) field theory on the brane's surface we have found a 6d moduli 
space for the scalar field as expected in the N=4 theory. 
Further allowing small velocities
in these directions, after a shift
in coordinate definitions in the $r- \theta$ plane, we find
\begin{equation}
L = {1 \over 2} \mu (\dot{r}^2 + r^2 d \Omega_5^2)
\end{equation}
which is a canonical kinetic term. This is to be compared to the N=4
kinetic term $\tau \Phi^\dagger \Phi$ from which we can see that 
the coupling is independent of $r$ or scalar vev. These two results
match with our expectations for the N=4 theory and provide evidence for the
correspondence. \vspace{-0.2cm}

\section{N=4 Broken to N=2}

A more interesting theory to study is one where we include a mass
term in the N=4 SYM theory that breaks the symmetry to N=2 in the IR
\cite{n2,PW,n21,probe}.
This theory should have a running coupling below the mass scale.
We must therefore allow an appropriate scalar to become non-zero
in the SUGRA, $m$. In addition one expects the N=2 
theory to have a moduli space parametrized by the vev $tr \phi_3^2$
of the remaining massless, complex scalar. In the SUGRA
this vev corresponds to the scalar $\rho$  discussed above in the N=4 theory.
There are again solutions of the SUGRA equations of motion where at
large y the scalar fields have vevs approaching zero as $exp(-dy)$
where d is the dimension of the source. In the IR these vevs grow.
The equations of motion allow solutions for any initial conditions
so we must understand how these correspond to physics in the dual
field theory. It is natural to fix the mass at some UV RG scale (value of y)
and then
look at solutions with varying $\rho$ (scalar vev) at that same scale
so the flows can be parametrized by $\rho(y_{UV})$.
As pointed out in \cite{n2} the flows divide into two classes,
those that fall down an unbounded potential and those that eventually
meet a rising potential. Inbetween there is a critical ``ridge'' flow.
Based on the analysis of such solutions at high temperature Gubser
has proposed that the flows that meet the rising potential are unphysical.
The remaining flows he identifies with different points on moduli space
with the critical flow perhaps representing some special point on
the N=2 theory's moduli space. Let us now look for confirmations
of these identifications.

The N=2 field theory has been studied in\cite{dorey}. 
There are indeed singular points
on the moduli space where monopoles become massless. When the theory 
is broken to N=1 by the inclusion of an extra mass $M$ for the massless
adjoint matter superfield of the N=2 theory a potential develops that pins
the vacua of the N=1 theory to these singular points. The mass $M$ can be 
introduced as a third scalar in the SUGRA\cite{n21} although now only
numerical solutions of the equations of motion are possible. Again sensible
boundary conditions are to fix $m$ and $M \ll m$ at some $y_{UV}$ and
the solutions are then parametrized by $\rho (y_{UV})$. In Fig 2 the
value of the potential at varying y values is shown
for each solution. Solutions to the right of the ridge flow still meet
a rising potential as they do in the N=2 solution. Now though solutions
to the left also begin to meet rising potentials in the IR. Eventually
only the ridge flow itself is still crossing a falling
potential. By Gubser's criteria only this single ridge solution should 
correspond to a physical vacuum of the field theory. In agreement 
with expectations it is the flow that was claimed to describe the
singular point on the moduli space.

\begin{figure}
\epsfxsize 12cm \centerline
{\epsffile{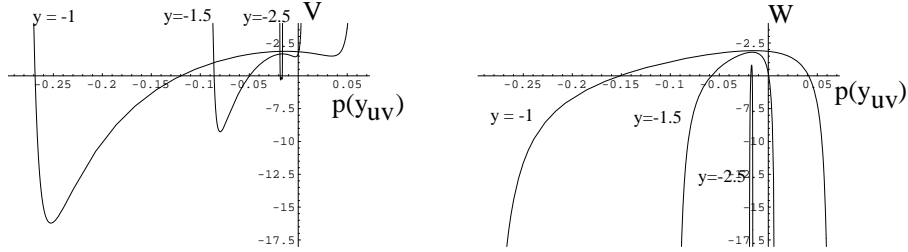}}
\caption{The potential and superpotential seen by 5d SUGRA flows 
in the N=4 broken to N=2 and then sequentially to N=1 theory, 
parametrized by $\rho(y_{UV})$,
as a function of radial coordinate.}
\end{figure}

We can also check the solution has the appropriate dimension moduli
space and running coupling by brane probing. To do so we again rely on
Pilch and Warner's ability to lift the 5d solutions. They provide us
with\cite{PW}
\begin{equation}
ds^2 = \Omega^2 ( e^{2  A} dx^2 + dr^2) + {\Omega^2 \over \rho^2}
( {d \theta^2 \over c} 
+ \rho^6 \cos^2 \theta  ( {\sigma_3^2 \over  c X_2} + 
{\sigma_1^2 + \sigma_2^2 \over X_1})+ {\sin^2 \theta
\over X_2} d \phi^2 )
\end{equation}
\begin{equation}
\Omega^2 = ( c   X_1 X_2 )^{1/4} / \rho, \hspace{1cm} c = \cosh m
\end{equation}
\begin{equation}
X_1 = \cos^2 \theta + \rho^6 c \sin^2 \theta, \hspace{1cm}
X_2 = c \cos^2 \theta + \rho^6  \sin^2 \theta
\end{equation}
\begin{equation}
C^4 = { e^{4A} X_1 \over \rho^2} dx^0 \wedge dx^1 \wedge dx^2 \wedge 
dx^3 
\end{equation}

Substituting into the Born Infeld action for a static probe D3 
brane\cite{probe,pol}
we find the potential only vanishes in the $r-\phi$ plane at $\theta = \pi/2$.
The moduli space is indeed two dimensional. Allowing motion in the 
radial direction in this plane we can find a radial 
coordinate $u$ in which the kinetic term is canconical 
\begin{equation}
L \sim  N {\cosh m  \over u^2 \sinh m}( \dot{u}^2  + u^2 \dot{\phi}^2)
\end{equation}

In these coordinates the $m$ behaviour decouples from that of $\rho$.
There are therefore a set of solutions, depending on the boundary
condition on $\rho$, that have $\rho$ diverge at different values
of $u$. We should cut off the flow of $m$ at this $u$ value since the SUGRA
diverges. 
Inserting the behaviour of $m$ as a function of radial coordinate in
the two asymptotic regimes we find that in the UV we return to the
N=4 value of the coupling and in the IR we find the form $N \log u/u_0$
where $u_0$ corresponds to the boundary condition on $m$.
The coupling therefore correctly interpolates between the N=4 beta function
and the N=2 IR beta function. It blows up at $u_0$. 
The flows with different $\rho(y_{UV})$
describe different points on moduli space and the low energy coupling is 
this function evaluated at the value of $u$ where $\rho$ diverges corresponding
to evaluating the beta function at the vev. Only for the critical flow 
discussed above does the divergence in $\rho$
fall within $u_0$ and here we must excise the solution at $u_0$ because  
the probe's tension has fallen to zero - this is the enhancon mechanism
of \cite{en}. It is indicative that in the field theory new
degrees of freedom become massless at these points on moduli space.
In \cite{pol} the authors point
out that to match the scalar kinetic term with the $F^2$ term's
coupling one must also make an angular change of variables that squashes
the enhancon ring at $u=u_0$ to a line segment. \vspace{-0.2cm}

\section{Summary}\vspace{-0.1cm}

We have seen therefore that the $AdS$/CFT continues to provide a good 
description of the gauge theory physics
when the N=4 SYM is deformed with relevant operators. This suggests that many 
other interesting aspects of strong interation gauge dynamics may be
open to a gravity description. One might even hope to eventually realize 
a gravity description of QCD. In the deformed theories one always suffers
from an inability to decouple the N=4 matter, since the N=4 theory is strongly 
coupled in the UV, but these ideas will hopefully lead to a wider class
of correspondences. Very recently this hope has begun to be realized as
a number of pure N=1 and N=2 theories have been shown to have gravity 
duals\cite{new}.  


\vspace{-0.2cm}



\begin{thebibliography}{99}


\bibitem{mald} J. Maldacena, Adv. Theor. Math. Phys. 2 (1998) 231,
hep-th/9711200.

\bibitem{review}O. Aharony, S. Gubser, J. Maldacena, H. Ooguri , Y. Oz, 
Phys. Rept. {\bf 323} (2000) 183, hep-th/9905111. 

\bibitem{RG}E. Witten, Adv. Theor. Math. Phys. 2 (1998) 253, hep-th/9802150.

\bibitem{def} L. Girardello, M. Petrini, M. Porrati and A. Zaffaroni,
JHEP {\bf 9812} (1998) 022, hep-th/9810126; 
J. Distler and F. Zamora, Adv. Theor. Math. Phys. 2 (1998)
1405, hep-th/9810206.

\bibitem{FGPW}D. Z. Freedman, S. S. Gubser, K. Pilch and
N. P. Warner, JHEP {\bf 0007} (2000) 038, hep-th/9906194

\bibitem{n2}S.S. Gubser, hep-th/0002160;
A. Brandhuber and K. Sfetsos, Phys. Lett. {\bf B488} (2000) 373, 
hep-th/0004148

\bibitem{PW} K. Pilch and N. P. Warner , hep-th/0004063.

\bibitem{n21}N. Evans, M. Petrini, Nucl. Phys. {\bf B592}
(2000) 129-142, hep-th/0006048 

\bibitem{gun} M. G{\"u}naydin, L.J. Romans and N.P. Warner, Nucl. Phys. B272
(1986) 598.

\bibitem{probe}N. Evans, C.V. Johnson, M. Petrini, JHEP {\bf 0010} (2000)
022,
hep-th/0008081.

\bibitem{dorey}N. Dorey  JHEP 9907 (1999) 021,  hep-th/9906011; 
N. Dorey and S. Prem Kumar,  JHEP 0002 (2000) 006,
 hep-th/0001103.

\bibitem{pol}A. Buchel, A.W. Peet, J. Polchinski,  hep-th/0008076 

\bibitem{en}C. V. Johnson, A. W. Peet and J. Polchinski,
Phys. Rev. D61 (2000) 086001, hep-th/9911161.

\bibitem{new}J. Maldecena and C. Nunez, hep-th/0007018; I.R. Klebanov, 
M.J. Strassler, JHEP 0008:052,2000, hep-th/0007191; M. Bertolini, P. 
Di Vecchia, M. Frau, A. Lerda,R. Marotta and I. Pesando, hep-th/0011077;
A. Rajaraman, hep-th/0011279.
\end{thebibliography}
\end{document}